\documentclass[12pt]{article}

\usepackage{times}
\usepackage{graphicx}
\usepackage{bm}

\begin{document}

\title{Monte Carlo study of shear-induced alignment of cylindrical micelles in thin films}
\author{Gaurav Arya and Athanassios Z. Panagiotopoulos \\
Department of Chemical Engineering \\
Princeton University \\
Princeton, NJ 08544, USA} \maketitle

\begin{abstract}
The behavior of confined cylindrical micelle-forming surfactants under the influence of shear has been investigated using Monte Carlo simulations.
The surfactants are modeled as coarse-grained lattice polymers, while the Monte Carlo shear flow is implemented with an externally imposed
potential energy field which induces a linear drag velocity on the surfactants. It is shown that in the absence of shear, cylindrical micelles
confined within a monolayer coarsen gradually with Monte Carlo ``time" $t$, the persistence length of the micelles scaling as $t^{0.24}$, in
agreement with the scaling obtained experimentally. Under the imposition of shear, the micelles within a monolayer align parallel to the direction
of shear, as observed experimentally. Micelles confined within thicker films also align parallel to each other with a hexagonal packing under
shear, but assume a finite tilt with respect to the velocity vector within the velocity-velocity gradient plane. We propose a novel mechanism for
this shear-induced alignment of micelles based on breaking up of micelles aligned perpendicular to shear and their reformation and subsequent
growth in the shear direction. It is observed that there exists a ``window" of shear rates within which such alignment occurs. A phenomenological
theory proposed to explain the above behavior is in good agreement with simulation results. A comparison of simulated and experimental
self-diffusivities yields a physical timescale for Monte Carlo moves, which enables an assessment of the physical shear rates employed in our
Monte Carlo simulations.
\end{abstract}

%\pacs{64.60.Cn, 81.16.Dn, 81.16.Rf, 82.35.Jk}

\section{INTRODUCTION}
The idea of ``driven" self-assembly  has gained widespread attention and application over the last few years. The concept involves using external
fields such as shear and electric fields to drive the molecular assembly process in a specific course, which may or may not be the direction taken
in the absence of the field. Of particular interest to us is the use of shear as a driving force to align self-assembled polymer microstructures
in the direction of the shear.

It is well-known that diblock or triblock copolymers self-assemble below an order-disorder transition temperature to form an array of striking
microphases, which usually take the form of spheres, cylinders, gyroids and lamellae \cite{Matsen96} in the bulk. Alignment of these diblock and
triblock copolymers microphases using shear is an old observation. Several authors \cite{Koppi93,Winey93,Kannan94,Chen97} have used oscillatory
shear during annealing to align diblock copolymer lamellae in the parallel or the transverse directions depending on the magnitude of the shear.
Others have demonstrated alignment of triblock copolymer cylinders in the shear direction whereby the cylinders pack themselves in a hexagonally
arrangement in the plane perpendicular to the shearing direction \cite{Keller70,Hadzii79a,Hadzii79b,Scott92,Winter93}.
 Several
mechanisms for the alignment of lamellae under shear have been proposed. Briefly, these include selective melting of unaligned portions of
microphases, annihilation of defects due to shear, and grain rotation due to rocking motion in the case of oscillatory shear \cite{Chen97}.
Substantially fewer studies have examined the alignment of micelles using shear \cite{Tepe95,Butler96,Yang00}. The mechanism proposed to explain
alignment in cylindrical domains has involved two steps. The first step requires breakup of cylinders perpendicular to shear and second step, the
subsequent rotation and alignment of shorter cylinders along the direction of shear \cite{Scott92}. Definitive experimental proofs and theories
relating to the above mechanisms of alignment are still lacking. More recently, a steady shearing of a {\em monolayer} of diblock cylinders under
annealing conditions has been shown to result in perfectly aligned diblock cylinders spanning distances as large as centimeters
\cite{Angelescu04}. Such alignment on surfaces (as opposed to bulk) is also very important from a technological point of view since the resulting
patterns have potential applications in lithography and semiconductor processing. It is not clear whether the alignment mechanism in these thin
films of polymers is similar to the mechanism hypothesized for the corresponding phenomena in bulk polymers. Clearly, a molecular-level
understanding of the shear-induced alignment mechanism is required.

The present study, motivated by the experimental findings of Ref. \cite{Angelescu04}, deals with understanding the molecular-level mechanisms
governing the alignment of thin films of cylindrical micelles under steady shear. The thin films to be examined in this study have thickness
ranging from one to five micelle diameters. We adopt a coarse-grained lattice Monte Carlo (MC) approach to model the surfactants, while the shear
is implemented by associating a ``pseudopotential" to it, which is then included within the MC acceptance criteria. Such an approach offered us an
opportunity to probe the relatively large lengthscales and long timescales involved in these phenomena, which are generally inaccessible using the
more atomistic approaches like molecular and brownian dynamics. Combined hydrodynamics-mean field approaches, though promising, fail to provide
molecular level information on the self-assembly process under shear \cite{Badalassi03}. Even brownian dynamics simulations of worm-like micelles
undergoing shear, which have been quite successful in modeling the micelles as a single polymer chain which can undergo scissions and
recombinations \cite{Kroger04}, provide little information on the actual self-assembly of individual surfactants which make up the micelles. In
the present study, our surfactants are modeled as shortened versions of diblock copolymers.

In this paper, we first investigate the coarsening behavior of micelles in annealing conditions in the absence of shear. It is shown that the
scalings obtained for the apparent kinetics of coarsening (in terms of MC ``time") match those obtained experimentally, thereby providing a good
basis for our subsequent studies. We then examine the effect of shear on the micelles. It is shown that the micelles align in the direction of
shear through a novel mechanism, similar to the one hypothesized by Scott et. al. \cite{Scott92}. The effect of film thickness on the alignment is
also examined. Finally, we examine the extent of alignment of micelles with respect to different shear rates ($\dot{\gamma}$) and temperature
($T$), and show a remarkable dependence of the quality of alignment on $\dot{\gamma}$ and $T$. We explain these trends quantitatively using a
simple model.

\section{MODEL}
Our model system consists of cylindrical micelles sandwiched between two impenetrable walls of comparable width. The special case of a monolayer
of these micelles is illustrated in Fig. 1a. Amphiphiles are modeled as an H$_{4}$T$_{4}$ lattice surfactant on a cubic lattice with coordination
number $z=26$, where H and T refer to the head and tail sites respectively \cite{Larson85}. The subscript in the above notation indicates the
respective lengths of the H and T units (refer to Fig. 1b) . Solvent molecules are assigned to occupy single sites and do not interact with the
amphiphiles. The amphiphile sites interact with each other through nearest neighbor and excluded volume interactions with interaction energies
$\epsilon_{HT}$, $\epsilon_{TT}$, and $\epsilon_{HH}$ for the three interactions. There is only one relevant energy parameter $\epsilon$ defined
as
\begin{equation}
\epsilon = \epsilon_{HT} - \frac{1}{2}\epsilon_{TT} - \frac{1}{2}
\epsilon_{HH}
\end{equation}
For the sake of convenience, we impose that $\epsilon_{TT}=-2$ and all other interactions are set to zero, as in Ref. \cite{Larson85}. The
dimensionless temperature is then set according to $\epsilon$ as $T^{*}=kT/|\epsilon |$, where $k$ is the Boltzmann constant. For the rest of this
paper, all the temperatures have been scaled as above but the superscript (*) has been omitted for convenience. The above model of surfactants,
though simple, captures most of the essential features of surfactant self-assembly \cite{Larson85,Floriano99}. The MC simulations are performed in
the canonical ensemble using reptation moves (slithering-snake algorithm \cite{Binder94}) which utilize configurational bias for computational
efficiency \cite{Frenkel92,dePablo93}.

Since, the H$_4$T$_4$ surfactant naturally forms spherical micelles at low volume fractions ($\phi$) of surfactants, we maintain the surfactants
at large volume fractions ($\phi=0.5$) where they form cylindrical micelles. Using still larger volume fractions leads to formation of gyroid and
lamellar phases. The simulation box in Fig. 1a was chosen to have a lateral dimension ($L$) much larger than the periodicity of the resulting
micelles, which is on the order of about 10 lattice units (lu). Choosing large lateral dimensions avoids many problems associated with finite-size
effects. The thickness of the film ($h$) was varied from 10 to 60 lattice units. Periodic boundary conditions are used in the two lateral
directions. Similar to the experiments in Ref. \cite{Angelescu04}, the temperature during annealing (with or without the shear) is kept above the
glass transition temperature ($T_{g}$) but below the order-disorder temperature ($T_{ODT}$).

A linear shear flow was imposed within the slit using the method proposed in Refs. \cite{Xu97,Milchev99}. The main idea of this approach is to
assign a {\em fictitious} potential energy gradient $\nabla U_{d}$ to the drag force experienced by each amphiphile site given by
\begin{equation}
\nabla U_{d} = \zeta v_{x}
\end{equation}
where $v_{x}$ is the non-zero component of the velocity in the $x$ direction and $\zeta$ is a friction coefficient. For a linearly varying shear
flow, the velocity profile of the solvent is given by $v_{x}=\dot{\gamma} y$ where $\dot{\gamma}$ is the shear rate and $y$ is the coordinate
perpendicular to the slit. Apart from the usual interaction energy contribution $\Delta U$, the MC reptation moves on the amphiphiles also include
an additional energy term from each of the sites due to contribution from shear given by
\begin{equation}
\Delta U_{d,i} = \int^{x_{2}}_{x_{1}} -\zeta \dot{\gamma} y dx  =
- \zeta \dot{\gamma} y_{avg}(x_{2}-x_{1})
\end{equation}
in going from $(x_{1},y_{1},z_{1})$ to $(x_{2},y_{2},z_{2})$, where the term $y_{avg}=(y_{1}+y_{2})/2$ ensures microscopic reversibility. The
final acceptance criteria for a MC move then becomes
\begin{equation}
P_{acc} = min\left(1,\exp \left[-(\Delta U + \sum_{i=1}^{N} \Delta
U_{d,i})/kT \right]\right)
\end{equation}
where the shear contribution has been summed over $N=8$ sites of
the amphiphile.

The acceptance criteria of Eq.~4 hence ensures that the displacements of the amphiphiles are always biased in the flow direction, and the ``flow
velocity" may be obtained as the ratio of the net displacement of molecules and the MC ``time" counted in terms of number of MC cycles (mcc).
Here, 1 mcc is defined as a cycle of attempted MC moves on all the surfactants within the simulation box. We have demonstrated that this shearing
approach provides a reasonable approximation for shear-flow dynamics for both monatomics and polyatomics as long as the MC moves are chosen
realistically \cite{Arya04}. We anticipate that the amphiphiles will exhibit reptational dynamics within the tightly bound micelles, and hence our
choice of performing MC reptation moves seems fully justified here.

\section{RESULTS AND DISCUSSION}

\subsection{Annealing without shear}
Here, we discuss the coarsening dynamics of a monolayer of micelles with ``time" (measured in MC cycles) during annealing in the absence of shear.
The simulations were conducted at $T=6$, and the simulation box is chosen to have dimensions of $L=120$ and $h=10$, such that the micelles form a
monolayer of cylinders within the thin film. We start with a completely disordered state of the surfactants i.e. $T>T_{ODT}$, and then lower the
temperature to $T_{g}<T<T_{ODT}$ and record the evolution of the system. The glass transition temperature of the surfactant thin films,
$T_{g}\simeq 3.5$, was obtained from self-diffusivity ($D_{s})$ vs. $T$ and internal energy ($U$) vs. $T$ data. The temperature at which the $\log
D_{s}-T$ curve exhibited an inflection point was designated as $T_{g}$. This value of $T_{g}$ also agreed very well with the temperature below
which $U$ remains constant with respect to $T$, signifying extremely slow dynamics. The order-disorder temperature was estimated to be equal to
$T_{ODT}\simeq 9.5$. This was obtained via the calculation of the average number of T beads neighboring the {\em end} H bead ($N_{neighb}$). An
inflection in $N_{neighb}$ versus $T$, where $N_{neighb}$ is shown to increase with $T$, was indicative of $T_{ODT}$ at which point the surfactant
become disordered.

Figures 2a-c shows snapshots of the simulations during this annealing process, beginning from the disordered system in Fig.~2a. Within a short
interval the micellar structures have taken the form of well-developed cylinders which intermingle with each other and have only slight
orientational correlation with each other and themselves due the many topological defects present in the system (Fig.~2b), resembling the
entangled state of worm-like micelles. At later stages these defects annihilate and reduce in numbers leading to larger correlations between
micelles as seen in Fig.~2c.

In order to quantify the degree of alignment and orientation of micelles, we have computed the order parameter of the micelles given by
\begin{equation}
\psi({\bf r}) = \exp[2i \theta ({\bf r})]
\end{equation}
where we have defined $\theta$ to be the angle subtended by the normal of the micellar axis and the $x$-axis. To calculate the persistence length
of the micelles, we have evaluated the correlation between order parameters at distances $|{\bf r}|$ apart as given by
$\left<\psi(\bf{r})\psi^{*}(0)\right>$. The persistence length $\xi$ is then determined by fitting an exponential $\exp (-|{\bf r}|/\xi)$ to the
correlation, as described in \cite{Harrison02}. Figure~3 shows this correlation computed at different times versus the distance $|{\bf r}|$. Note
that all correlations decay to zero at finite distances, and more importantly, the correlations persist for longer distances at larger times. The
inset shows that the correlation lengths $\xi$ increase ever slowly with time with a scaling of $t^{0.24}$. A similar scaling is obtained when
$\rho^{-1/2}$ is plotted versus time (not shown), where $\rho$ is the number density of defects which is generally proportional to $\xi^{-2}$.

The above scaling agrees very well with the scaling of $t^{1/4}$ obtained experimentally by Harrison~et~al. \cite{Harrison02}. These authors
suggest a coarsening process dominated by a defect annihilation mechanism to be the origin for the above particular scaling, which agrees well
with our observations. Hence, our simple lattice polymer model is able to reproduce the complex coarsening mechanism obtained experimentally,
which provides support for the validity for our subsequent results using shear with this highly simplified model.

\subsection{Shear-aligning mechanism}
In order to study the effect of shear, the monolayer system with $h=10$ described above is now sheared while annealing it from the disordered
system shown in Fig.~2d. A multilayer version ($h$ upto 60) is discussed in the next section. Figure~2e shows a snapshot of the sample after a
short time whereby some of the micelles have aligned in the direction of shear. Shearing the sample for longer times yields perfectly aligned
micelles shown in Fig.~2f, similar to the flawlessly aligned diblock cylinders in Ref. \cite{Angelescu04}. We have established a simple mechanism
for alignment by analyzing snapshots of the system at intermediate stages. Starting from a disordered state, micellar structures begin to appear
and coarsen with time. At the same time, micelles also elongate by merging with nearby micelles and isolated amphiphiles. It is important to note
that the micelles tend to grow in the direction of shear due to an effective increase in the transport (diffusion) rates parallel to the shear as
opposed to the direction perpendicular to it. This has been verified through calculation of diffusion coefficients in the two directions. Micelles
oriented perpendicular to the shear direction either break apart when they become too long, or merge with other micelles in the direction of shear
when they are short.

The mechanism of shear alignment is even more vividly illustrated by shearing a pre-aligned micellar system (shown in Fig.~4a) in a direction
perpendicular to its original alignment. Due to increased transport rates in the shear direction, local instabilities are created within the
micelles which result in individual micelles to approach each other at intermittent locations as shown in Fig.~4b. This causes the micelles to
break up into smaller pieces as shown in Fig.~4c. Due to the shear, these broken micelles extend and join with existing micelles in the direction
of shear, thereby forming ``bridges" as shown in Fig.~4d. The originally intact micelles now break apart between the bridges to give rise to
structures which are now roughly oriented in the direction of shear as shown in Fig.~4e. The above process occurs within a very short time. These
roughly aligned micelles now take a comparatively longer time to attain the almost perfectly aligned state shown in Fig.~4f. The orientational
correlation function for the final shear-aligned sample does not decay to zero within the bounds of the simulation box (not shown). The cylinder
diameter and the spacing between the micellar cylinders also decreases sometimes with prolonged shearing, as observed in some systems
experimentally \cite{Jackson95}. It was also observed that he above described mechanism applied equally well for the shear-induced alignment of
multilayered cylindrical micelles. Note that such a mechanism for alignment agrees very well with that hypothesized by Scott et~al. \cite{Scott92}
for the case of alignment in triblock copolymer cylinders discussed before. The effect of shear rate and temperature on alignment is discussed
later.

We also computed the orientation angle of individual amphiphiles within the state of alignment and found that the molecules prefer to maintain a
slight tilt with respect to the micellar axis in the direction of shear, with angles ranging from 90 to 85$^\circ$ as the shear rate is increased.
This may be one of reasons for the reduction in diameters of the micelles with shear rate observed in the simulations.

\subsection{Effect of film thickness}
So far, we have only considered shearing of a monolayer-thick surfactant film. We now explore the impact of the film thickness on the resulting
structure of micelles after shearing. To this end, we have conducted shear MC simulations at several film thickness ranging from $h=10$ to $h=60$
lattice units. The length of the simulations were chosen to be long enough to reach steady state. Due to the fact that the computational
requirements of the MC simulations increase drastically with the film thickness, we have restricted our analysis to only a few micelles-thick
films. The main results from this study are summarized in Figs.~5 and 6.

Figure~5 shows the isometric snapshots of the resulting micelles at different film thickness after shearing them for 1 million MC cycles. The
shear rates for each film thickness in Fig.~5 have been chosen to be approximately 20-30{\%} below the critical shear rate at which the systems
become disordered. Examining the values of shear rates mentioned in Fig.~5, it may be concluded that as the films become thicker, they become more
prone to disorder with respect to the magnitude of the imposed shear rate. The reason for this is discussed in more detail in the next section. As
discussed before, the micelles corresponding to $h=10$ adopt a parallel orientation to the velocity vector (${\bf v}$). The micelle diameter
($d\simeq 11$) is slightly smaller than the film thickness, which gives their crosssection a slight elliptical appearance (see Fig.~5a). A thicker
film with $h=20$ is not able to accommodate a bilayer of micelles with the dimensions mentioned above, which results in the formation of a single
layer of lamellar micelles. As the thickness is further increased to $h=30$, the walls are able to accommodate a bilayer of cylindrical micelles.
The micelles align parallel to each other as in the monolayer case. However, they now assume tilt of approximately $28^{\circ}$ with respect to
the velocity vector within the shearing plane (${\bf v}$-${\bf \nabla v}$), while still remaining parallel within the velocity-vorticity plane
(${\bf v}$-${\bf e}$). Longer simulations still yielded the same tilt angle. As, the thickness is further increased to $h=40$, the tilt angle
decreases to about $18^{\circ}$, as seen in Fig.~5. In addition, the micellar cylinders are seen to pack themselves in a hexagonal fashion, with
the [10] plane parallel to the vorticity direction, as seen experimentally by Tepe et al. \cite{Tepe95}. A further increase in the thickness only
causes a minor decrement in the tilt angle (refer to Fig.~6). Finally, in order to demonstrate that the tilt phenomenon is real and not an
artifact of our choice of the shear rate or due to finite-size effects, we conducted simulations at different shear rates and system sizes. It was
observed that the shear rate as well as the system size had an insignificant effect on the tilt angle. The results for the latter case are
summarized in the inset of Fig.~6. Hence, it may be reasonable to conclude that the tilt observed in the micelles is a real phenomena.
Extrapolating the tilt angles in Fig.~6 to much thicker films suggests that the micelles would still assume a small but finite tilt angle in the
bulk. More extensive simulations are needed to prove this, which are beyond the scope of this paper.

The observed tilt in the micelles contradicts the results of several experiments conducted in the bulk, where the block copolymer cylinders have
been demonstrated to nearly align parallel to the velocity vector, both under shearing and extensional flows
\cite{Scott92,Winter93,Tepe95,Jackson95}. In these instances, the block copolymer domains were essentially ``infinitely" long. On the other hand,
our results agree with the extensional flow experiments of Pakula et al. \cite{Pakula85} who observed that short copolymer domains are inclined at
a finite angle ($\sim 20^{\circ}$) to the velocity vector. Theoretical studies as well as nonequilibrium molecular dynamics simulations of long
hydrocarbons \cite{Kioupis99} and worm-like micelles (modelled as polymers with variable length) \cite{Kroger04} in shear flow also confirm a
tilting of the molecular entities with respect to the shearing direction. From the above discussion, we may {\em hypothesize} that the length of
the micelle is an important factor in governing its subsequent alignment under shear. Long micelles tend to adopt a parallel orientation, while
short ones tend to assume an inclination to the shearing direction. This difference is possibly due to the fact that the lengthy micelles
experience significantly larger torques due to shear as compared to the short micelles. A more extensive analysis of this phenomena as well as its
experimental verification is required.

\subsection{Effect of shear rate and temperature}
It is important to quantify the effect of different shear rates and temperatures on the rate as well as the quality of alignment of micelles. To
this end, we have performed simulations at various values of the quantity $\zeta\dot{\gamma}$, which characterizes the magnitude of the resulting
shear rate, and temperatures for the $h=10$ system. The analysis performed here can be readily extended to the case of thicker films. The results
are summarized in Fig.~7. It is observed that there exist three distinct regimes. In the first regime, the shear rates are too low to influence
any alignment to the micelles within the time frame of our simulations ($\sim$ 1.5 million MC cycles). The micelles simply coarsen with time
without any major changes in their orientations. Since alignment requires micelles to break up into smaller chunks according to the mechanism
outlined before, the shear should be sufficiently strong in order to provide the necessary energy for micellar breakup. In the second regime the
shear is strong enough to influence the micelles to align within reasonable amounts of simulation time. When the shear becomes too strong, it
results in micelles breaking up irreversibly resulting in a completely disordered system. This is the third regime which in many ways is similar
to the disordered states obtained for $T>T_{ODT}$. The extent of the three regimes with respect to the shear rate obeys an interesting temperature
dependence which may be quantified by a simple phenomenological theory described next.

The lower bound on the shear rate ($\zeta\dot{\gamma}_{lower}$) which results in alignment may be estimated by noting that such alignment would
occur only when the shear energy term in the acceptance criteria (Eq.~1) is of the same order of magnitude as the energetic interactions term
alongside it. Since the interaction energy term is proportional to $|\epsilon |$ and that the shear energy term is proportional to
$\zeta\dot{\gamma}$, we may then assert that
\begin{equation}
\zeta\dot{\gamma}_{lower} = a |\epsilon |
\end{equation}
is valid to the first order. Note that the constant $a$ is roughly inversely proportional to the height of the film ($h$), which evidently arises
from the term $y_{avg}$ within the shear energy term in Eq.~3. Also, $a$ is clearly dependent on the micelle architecture. It may be assumed for
convenience here that $a$ is independent of temperature. Equation~6 along with a fit parameter $a=0.24$ seems to agree very well with the
simulation results as shown in Fig.~7. The upper bound on the shear rate refers to the critical value of shear rate where the sum of the shear
energy term, which is proportional to $\zeta\dot{\gamma}$, and the interaction energy becomes equal to the energy of disordering, the latter being
proportional to $T_{ODT}$. This requires that
\begin{equation}
\zeta\dot{\gamma}_{upper} = b (T_{ODT}-T)
\end{equation}
where $b$ is another constant similar in nature to $a$ i.e. it is expected to be inversely proportional to $h$ and dependent on the surfactant
architecture. Again, the upper bound with $b=0.31$ agrees very well with the simulation results. These results suggest that the shear aligning
regime is fairly broad at low temperatures but diminishes linearly with increasing temperature. At temperatures higher than the temperatures where
the two bounds intersect ($T\sim 8.6$), it is impossible to attain shear alignment. Also note that shear aligning does not continue indefinitely
as the temperature is decreased because the system reaches a glassy state whereby the dynamics of the system become so slow that alignment becomes
impossible to attain.

An increase in the film thickness ($h$) is expected to shift the lower bound downwards and decrease the slope of the upper bound as a result of
the inverse dependence of $a$ and $b$ on $h$, as discussed before. Nevertheless, the qualitative nature of the phase diagram is expected to remain
unchanged. The above dependence of $b$ on $h$ explains the reason why progressively small shear rates were used to align micelles in Fig.~5 as the
films became thicker.

It is also instructive to describe the kinetics of shear-induced alignment with respect to the magnitude of the shear. Figure~8 shows the nematic
order parameter $\left< 2\cos^{2} \theta -1 \right>$ plotted against Monte Carlo time for $T=6$ at different shear rates. It can be clearly seen
that the initial rate of alignment (for $t<10000$ MC cycles) increases monotonically with the shear rate. However, the long-term quality of
alignment, which is reflected by the limiting value of the order parameter, increases initially with the shear rate whereby it reaches its
maximum, and then decreases with the shear rates as disordering sets in at large shear rates. Based on the above results, we propose that the most
efficient procedure for obtaining well-aligned micelles is to begin the annealing process at a large shear rate (but still maintaining
$T<T_{ODT}$) during which the micelles align fairly rapidly with time and then gradually bring down the shear rate so to improve the final quality
of alignment as evident from Fig.~8.

\subsection{MC versus experimental shear rate}
It is instructive to obtain a rough estimate of the shear rates employed in our MC simulations to align the micelles, and compare them with the
actual shear rates used in the experiments of Ref. \cite{Angelescu04}. To this end, we need to match the length and timescales of MC simulations
to physical units of meters and seconds, respectively. The protocol consists of first obtaining an estimate of the real length of our lattice
surfactant, and then estimating the size of an MC cycle by comparing the translational diffusivity ($D_{s}$) obtained through MC simulations with
that from experiments, under comparable conditions. The shear rate obtained from simulations may then be quickly translated into real units of
s$^{-1}$. It should be kept in mind that any such matching where a lattice model is mapped onto its off-lattice and flexible version is bound to
be crude and hence would only be accurate to within an order of magnitude.

Through a comparison of the solubility data of hydrocarbons in water at a temperature of 298 K (obtained by Tsonopoulos \cite{Tsono99}) with that
computed for our lattice polymers from their phase behavior, it may established that a single T bead corresponds to roughly an ethyl unit
(C$_{2}$) and a single H bead corresponds to an ethoxy group (E$_{1}$), while a vacant solvent site corresponds to two water molecules
\cite{Panag}. Hence, our H$_4$T$_4$ surfactant corresponds to the real surfactant C$_8$E$_4$, and the lattice unit spacing in our simulations
amounts to a length of roughly 3 {\AA} i.e $1$~lu $\simeq 3 \times 10^{-10}$~m. It may also be established through such a comparison that
$\epsilon \simeq 60$~K. Accordingly, our simulations conducted at $T=6$ correspond to a real temperature of 360~K. We have also calculated the
self-diffusivity of our surfactants in the cylindrical phase at $\phi=0.5$ and $T=6$ to be approximately equal to $D_{s}=3.4 \times
10^{-6}$~lu$^{2}$/mcc. From experimental studies on C$_{8}$E$_{5}$ surfactants \cite{DErrico02} and C$_{8}$E$_{4}$ surfactants \cite{Frindi92}, as
well as molecular dynamics studies on model surfactants C$_{19}$ \cite{Maiti02} under conditions of temperatures and volume fractions similar to
our simulations, it may be established that the the C$_{8}$E$_{4}$ surfactant has a self-diffusivity on the order of $10^{-10}$~m$^{2}$/s. This
sets a physical time of $3.1 \times 10^{-15}$~s for our MC cycle (1 mcc) via comparison of the simulated and experimental diffusivities.

The MC shear rates measured in our simulations at $T=6$, corresponding to the shear aligning regime $\zeta\dot{\gamma} \in (0.24,1.09)$ in Fig.~6,
ranged from $1.3 \times 10^{-6}$/mcc to $5.7 \times 10^{-5}$/mcc. Using the physical time associated with each MC cycle, these translate to
physical shear rates ranging from $4 \times 10^{8}$/s (denoted by $\dot{\gamma}_{L}$) to $1.8 \times 10^{10}$/s. The length of the MC simulation
runs in this study, which were several million mcc, correspond to merely a few nanoseconds of real time. On the other hand, the experimentally
employed shear rates needed to align the diblock copolymers in Ref. \cite{Angelescu04} were on the order of 10/s and the alignment occurred over
several hours. At first glance, it may seem that the conclusion drawn from our previous analysis, namely that shear rates lower than
$\dot{\gamma}_{L}$ do not result in any alignment of the micelles, is violated by the experimental evidence of alignment at small shear rates of
10/s. It should however be pointed out that our analysis only considered alignment within the first few nanoseconds while the experiments span a
time of several hours. Clearly, observing the samples over much larger times may show evidence of alignment at shear rates equal to, and much
smaller than, $\dot{\gamma}_{L}$. We therefore believe that the validity of Eq.~(6) would hold, though the lower bound on the shear rate
characterized by the constant $a$ in Eq.~6 is expected to decrease, with a severalfold increase in the ``observation" time. Such drastic
differences in the accessible shear rates and timescales clearly point to the apparent limitations of molecular simulations.

\section{CONCLUSIONS}

We have studied the behavior of self-assembling, cylinder-forming surfactants, both in the presence and absence of shear. It is shown that in the
absence of shear, the cylindrical micelles coarsen via a defect annihilation mechanism with a kinetic scaling of $t^{0.24}$, very similar to the
mechanism and scalings observed experimentally. In the presence of shear, it is observed that the micelles align rapidly with shear through a
mechanism where the micelles perpendicular to the shearing direction break up into smaller micelles, and that these smaller micelles align and
grow in the direction of shear via recombination with other micelles and individual surfactant chains. This mechanism, responsible for the
alignment of micelles, agrees very well with mechanisms previously proposed for the alignment of triblock copolymer cylinders. It is shown that
the micelles align perfectly parallel to the shearing direction for monolayer-thick films, while they assume tilts of 15-30$^{\circ}$ with respect
to the shearing direction in the shearing plane for thicker films which can accommodate multilayered micelles. We hypothesize that this is as a
result of the micelles being fairly short. It is demonstrated that there exists a window of shear rates within which shear aligning takes place,
and that this window decreases linearly with temperature. We have also developed a simple theory for predicting this domain of shear alignment on
a temperature and shear rate plane, which agrees well with our simulations. At the end, we have performed a comparison of the shear rates utilized
in our study with those used experimentally to align diblock copolymers. We conclude this study by stating that coarse-grained lattice models of
polymers combined with novel dynamical features offer a powerful approach to understanding phenomena occurring in the sub-continuum regime.

\section*{ACKNOWLEDGMENTS}
This work was supported by the NSF (MRSEC Program) through the Princeton Center for Complex Materials (DMR 0213706). Additional support has been
provided by ACS-PRF (grant 38165-AC9). We also thank Paul Chaikin and Rick Register for useful discussions.

\begin{figure}[t]
\includegraphics[angle=0,height=3.2in]{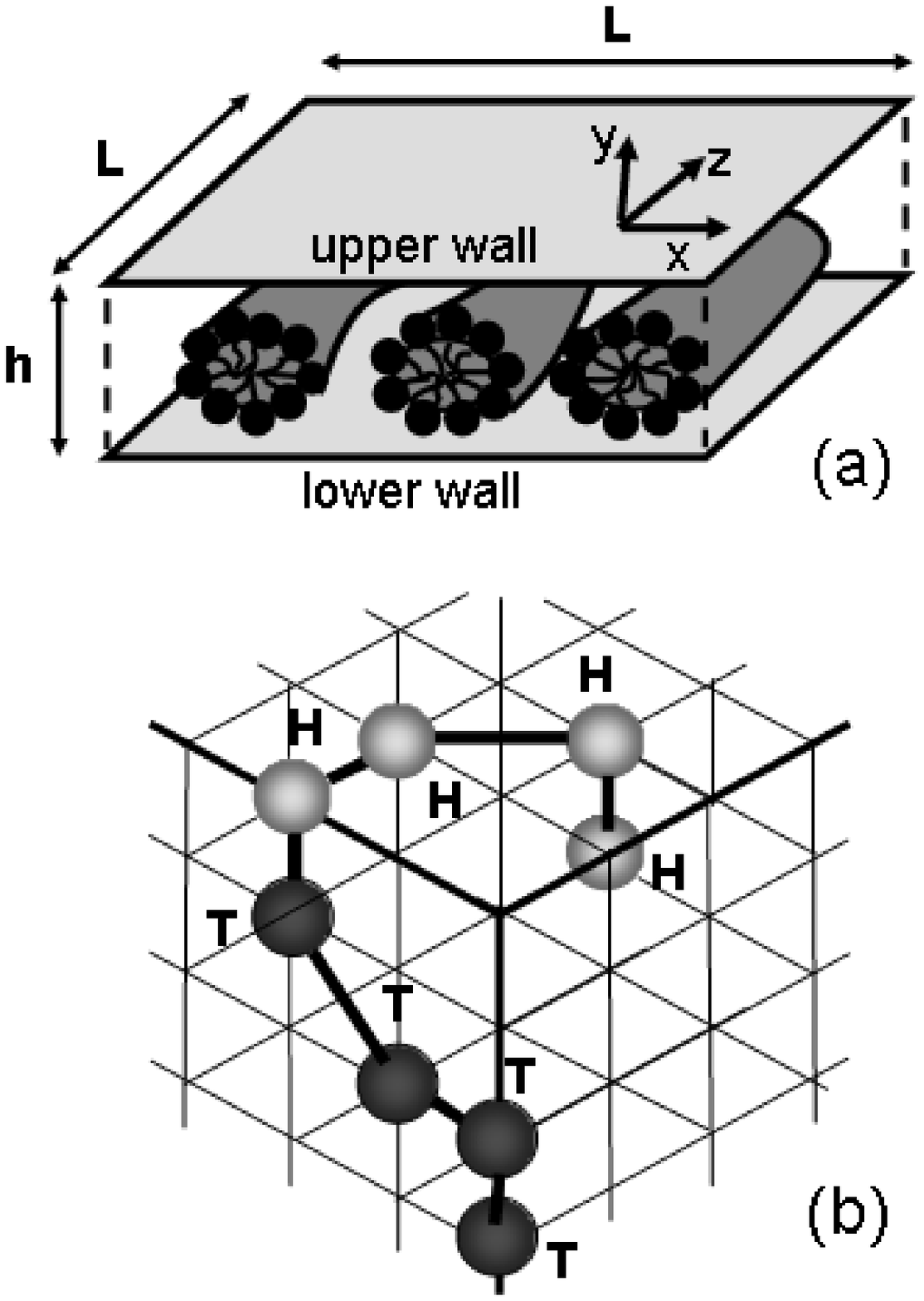}
\caption{\label{fig1} a) Schematic of our model system showing a one-dimensional layer of cylindrical micelles confined in a slit pore, and  b) a
cartoon of an H$_{4}$T$_{4}$ amphiphile on a cubic latice.}
\end{figure}

\begin{figure*}[t]
\includegraphics[angle=0,width=4.5in]{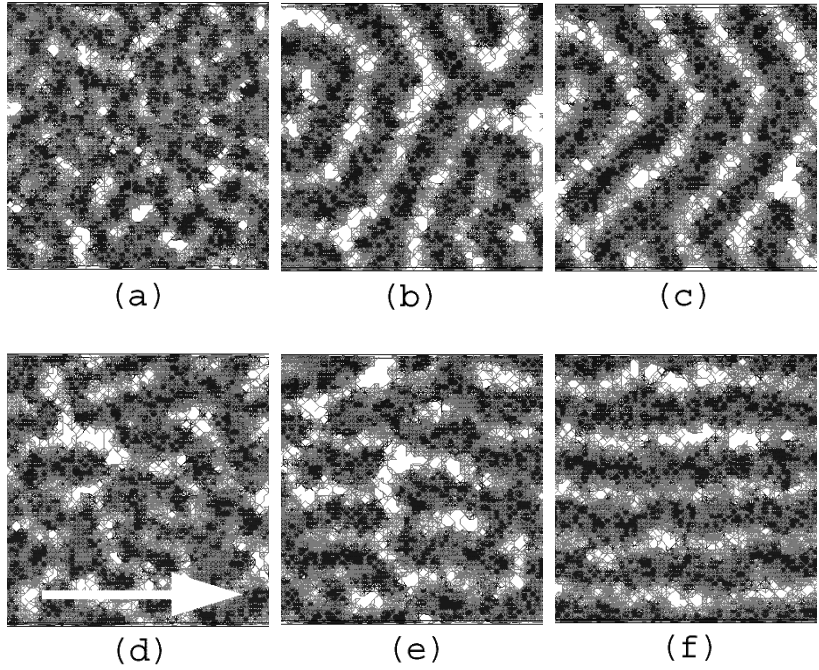}
\caption{\label{fig2}  Top view of the micellar system at different stages during annealing at $T=6$. Snapshots (a)-(c) show coarsening of
micelles during annealing without shear at (a) $t=0$, (b) $t=0.1$, and (c) $t=3.5$ million MC cycles. Snapshots (d)-(f) show the alignment of
micelles under a shear rate of $\zeta\dot{\gamma}=0.72$ in the direction of the arrow at d) $t=0$, e) $t=0.28$, and f) $t=1.4$ million MC cycles.
All snapshots show a $60 \times 60$ lu$^{2}$ region within a simulation box of dimensions $L=120$ lattice units. The black and grey colored
regions refer to the tail and head sites respectively.}
\end{figure*}

\begin{figure}[t]
\includegraphics[angle=90,width=3.5in]{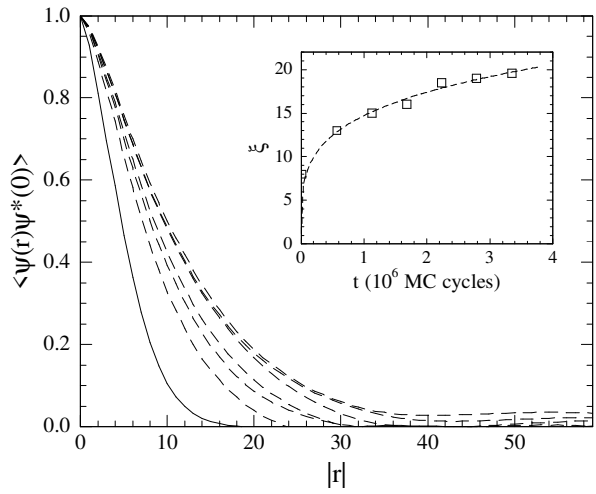}
\caption{\label{fig3} Orientational correlation between micelles as a function of distance during annealing. The solid line represents the
correlation at $t=0$ while the dashed lines represent times of 0.57, 1.13, 1.68, 2.24, 2.79 and 3.35 million MC cycles progressing outwards from
the origin. The inset shows the corresponding correlation lengths plotted as a function of the annealing time. The dashed line in the inset
corresponds to a power-law fit of $t^{0.24}$.}
\end{figure}

\begin{figure*}[t]
\includegraphics[angle=0,width=4.5in]{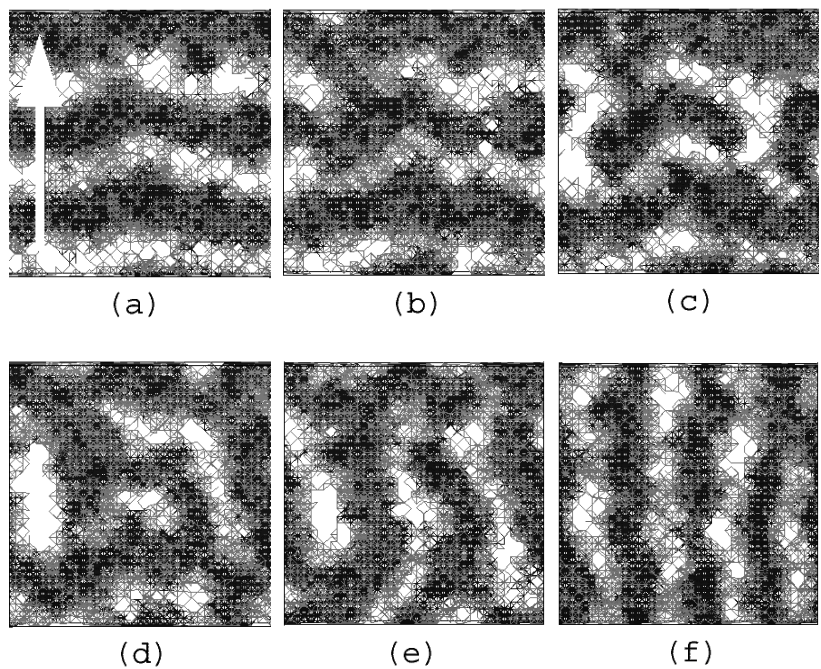}
\caption{\label{fig4} Snapshots of a small simulation box with $L=40$ lattice units showing the evolution of a pre-aligned micellar system in the
presence of a shear in a direction perpendicular to the initial alignment of micelles at (a) $t=0$, (b) $t=0.005$, (c) $t=0.0125$, (d) $t=0.025$,
(e) $t=0.0325$, and (f) $t=0.25$ million MC cycles, respectively.  The shading format and significance of arrow  is as in Fig.~2.}
\end{figure*}

\begin{figure*}[t]
\includegraphics[angle=90,width=5.4in]{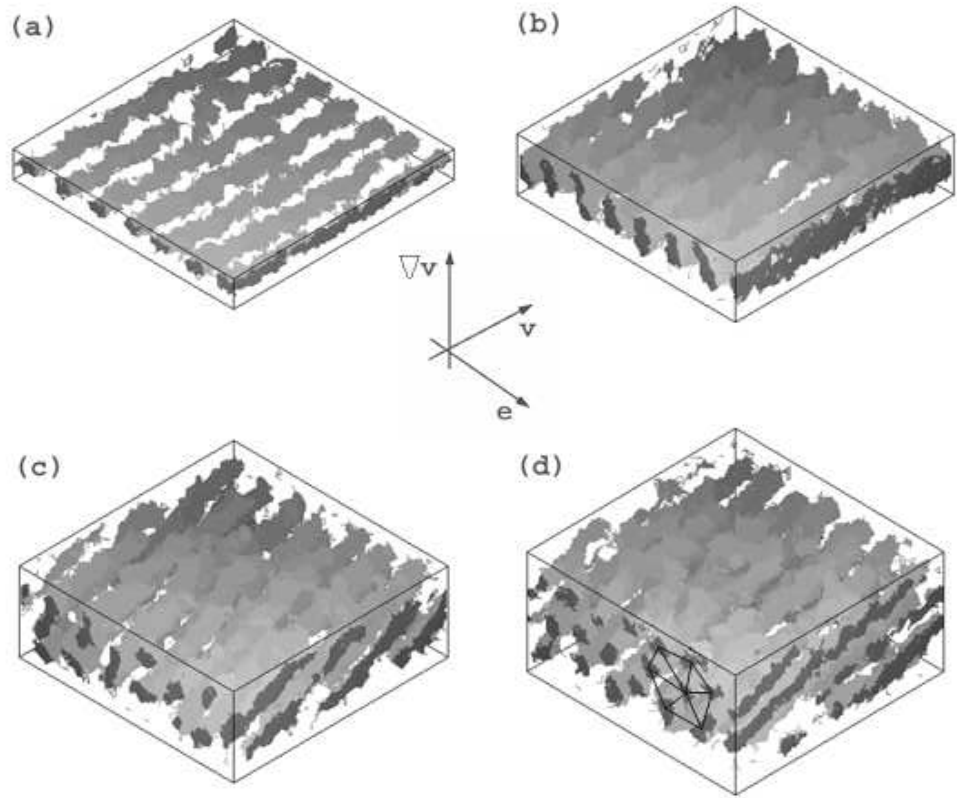}
\caption{\label{fig5} Isometric view of the surfactant thin films after annealing them under shear at $T=6$ for 1 million MC cycles for (a) $h=10$
and $\zeta\dot{\gamma}=0.72$, (b) $h=20$ and $\zeta\dot{\gamma}=0.36$, (c) $h=30$ and $\zeta\dot{\gamma}=0.18$, and (d) $h=40$ and
$\zeta\dot{\gamma}=0.12$. The corona of micelles composed of T beads has been ommited for clarity. The light grey-shaded region corresponds to the
core of the micelles while the dark-grey regions represent to the crosssection of micelles along the faces of the simulation box. The hexagonal
packing of cylinders in (d) has been emphasized by connecting cylinder centers via black lines.}
\end{figure*}

\begin{figure}[t]
\includegraphics[angle=90,width=3.5in]{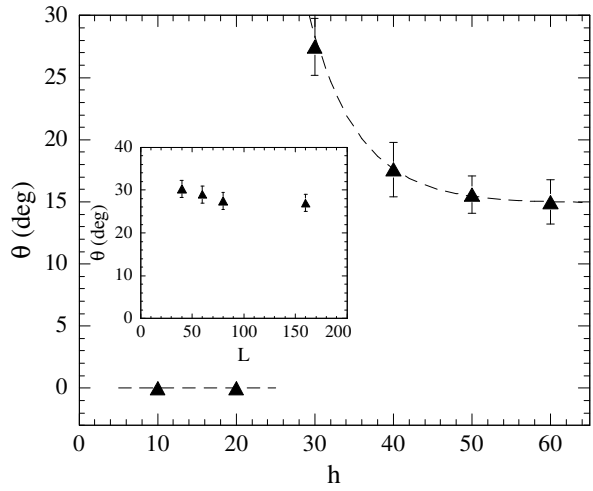}
\caption{\label{fig6} Dependence of the alignment angle on the film thickness. The symbols show the simulation data while the dashed lines are
meant to guide the eye. The inset shows the alignment angle as a function of lateral system size.}
\end{figure}

\begin{figure}[t]
\includegraphics[angle=90,width=3.5in]{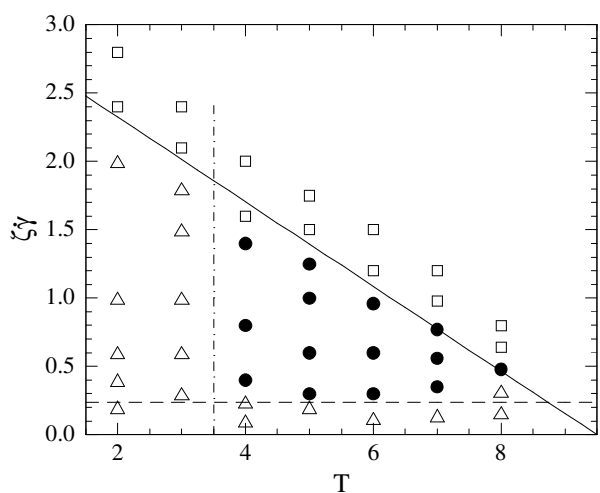}
\caption{\label{fig7} Pseudo-phase diagram showing regions of shear alignment as a function of temperature and shear rate. The symbols represent
simulation results showing regions of shear alignment (solid circles), regions of well-formed but unaligned micelles (triangles), and regions of
disorder (squares). The solid and dashed lines show the analytical upper and lower bounds of regions where shear aligning occurs while the region
left of the dash-dotted line represents the glassy region where shear alignment does not occur.}
\end{figure}

\begin{figure}[t]
\includegraphics[angle=90,width=3.5in]{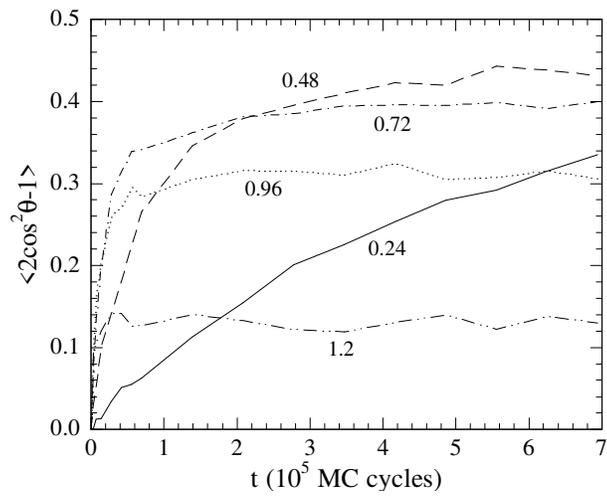}
\caption{\label{fig8} Evolution of the nematic order parameter with time at different shear rates at $T=6$. The curves have been labelled with
their corresponding shear rate.}
\end{figure}

\end{document}